\documentclass[reprint,amsmath,amssymb]{revtex4-1}
\usepackage[utf8x]{inputenc}
\usepackage{graphicx}
\usepackage{dcolumn}
\usepackage{bm}
\usepackage{color} 
\usepackage{amsmath}
\usepackage{tensor}

\usepackage{soul} 
\usepackage[normalem]{ulem} 

\begin{document}

\preprint{APS/123-QED}

\title{Topological indicators for systems with open boundaries: Application to the Kitaev wire}

\author{Bal\'azs Het\'enyi$^{1,2}$, Andr\'{a}s L\'{a}szl\'{o}ffy$^{1}$, Karlo Penc$^{1}$, and Bal\'{a}zs \'{U}jfalussy$^{1}$}
\affiliation{$^{1}$Institute for Solid State Physics and Optics, Wigner Research Centre for Physics,  H-1525 Budapest, P. O. Box 49, Hungary\\}
\affiliation{$^{2}$MTA-BME Quantum Dynamics and Correlations Research Group, Department of Physics, Budapest University of Technology and
  Economics, H-1111 Budapest, Hungary \\}

\date{\today}
\begin{abstract}
\textcolor{black}{
To clarify the relationship between edge electronic states in open-boundary crystalline systems and their corresponding bulk electronic structure, Alase et al. [Phys. Rev. Lett. 117, 076804 (2016)] have recently generalized Bloch's theorem to lattice models with broken translational symmetry.  Their formalism provides a bulk-boundary correspondence indicator, D, sensitive directly to the localization of edge states.  We extend this formalism in two significant ways.  First, we explicitly classify the edge-state basis provided by the theory according to their underlying protecting symmetries.  Second, acknowledging that the true topological invariant is the Zak phase, inherently defined only for periodic boundary conditions, we introduce an analogous quantity $\gamma_Z$, suitable for open-boundary systems.  We illustrate these developments on the example of the Kitaev wire model.  We demonstrate that both indicators, $D$ and our open-boundary analog, $\gamma_Z$, effectively capture topological localization: while D diverges within the topological phase, $\gamma_Z = \pi$ .  We further examine eigenvalues ($z$) of the lattice shift operator, showing that variations in these eigenvalues as a function of chemical potential provide additional signatures of topological phase transitions.  Additionally, we study a periodic Kitaev wire containing a bond impurity, revealing that while the topological phase remains characterized by  $\gamma_Z = \pi$, significant state localization near the impurity emerges only at critical values of the chemical potential.”}
\end{abstract}

\maketitle


\section{Introduction}

The relation between edge electronic states of an open boundary crystalline system and the bulk electronic structure is a question with a long history, which received impetus recently with the development of topological insulators~\cite{Bernevig13,Asboth16} and topological superconductors~\cite{Qi11,Sato17}.  Such systems exhibit phases which are characterized by a topological invariant, a number defined for a given model under periodic boundary conditions (PBC).  In accordance with the bulk-boundary correspondence (BBC) principle, when the topological invariant in the PBC version of a given model takes a nontrivial value, the version of the same model with open boundary conditions (OBC) exhibits edge states at zero energy.   \textcolor{black}{Recently a generalization of the BBC principle was also developed, the spectral BBC principle~\cite{Tamura19,Daido19,Tamura21,Ahmed25}, which relates frequency dependent quantities between the edges of an OBC system to those of the bulk of a PBC version of the given model.}  The nature of these edge states can vary widely.  In the Su-Schrieffer-Heeger model~\cite{Su79} it is a localized boundary mode.  In two dimensional, time-reversal invariant topological insulators~\cite{Kane05a,Kane05b}, the boundary states carry a spin current.  In the Kitaev $p$-wave superconducting wire~\cite{Kitaev01}, the edge mode is a fractionalized quasi-particle excitation of the Majorana fermion type.   Topological edge states also occur in non-Hermitian systems~\cite{Torres19,Ashida20,Bergholtz21,Zhang22,Okuma23,Yao18,Yokomizo19,Yang20,Verma24}.  The effect of impurities~\cite{Muller16,Tinyukova19,Mishra21,Woods21,Peeters24} and disorder~\cite{Akhmerov11,Brouwer11a,Brouwer11b,DeGottardi13,Sau13,Adagideli14} in Majorana fermions is also a crucial question from the point of view of topological quantum computing.  \\

\textcolor{black}{According to Kitaev's original suggestion, the setup for realizing the Majorana quasiparticle was a semiconducting wire situated on the surface of a three-dimensional $p$-wave superconductor.  Roughly speaking, the Majorana quasiparticle arises as a linear combination of half a particle-like and half a hole-like contribution as zero energy modes located at the ends of the wire.  Kitaev also demonstrated~\cite{Kitaev01} the stability of the Majorana fermion to decoherence which makes it an ideal platform for constructing quantum bits for topological quantum computation~\cite{Alicea12,DasSarma15}.   The realization of the Kitaev wire is not a trivial feat, but there has been much progress in recent years~\cite{Kitaev01,Alicea12,DasSarma15,Lutchyn18,Zhang19,Frolov20,Beenakker20,Marra22,Nyari23,Laszloffy23,Tanaka24,Rachel25}.   The three main suggested platforms of these research efforts are: a superconductor junction on the surface of a strong topological insulator~\cite{Fu08}, semiconductor heterostructures with strong spin-orbit coupling~\cite{Sau10}, and magnetic chains on the surface of a superconductor~\cite{Nadj-Perge13}.  Recent studies of some of the authors of this work focused on a finite chain of magnetic atoms on a superconducting substrate (for example, niobium), possibly with an epitaxial layer of gold (to increase spin-orbit coupling).  First principles calculations for these types of systems with artificially constrained spin spiral configurations show evidence for Majorana fermions at the edges of the atomic chain~\cite{Nyari23,Laszloffy23}.}\\

\textcolor{black}{Theoretical efforts~\cite{Tamura19,Daido19,Tamura21,Ahmed25,Gurarie11,Sticlet12,Fulga12,Sedlmayr15,Peng17,Katsura18,Awoga24,Karoliya25,Alase16,Alase17,Cobanera17,Cobanera18} have focused on the unambiguous identification of topological edge states in general, and in particular of the Majorana fermion.   Generally speaking, minimalist models (such as the Kitaev wire) allow for the direct calculation of the topological invariant for a PBC system, and the detection of localized edge states in the OBC system, however, the situation is often more challenging for more realistic models.    The presence of localized edge states~\cite{Davison92} is not necessarily indicative of a Majorana fermion.  Orbitals corresponding to dangling bonds, or the presence of disorder~\cite{Abrahams79,Langedijk09} can all give rise to localized edge states.  In the case of magnetic atoms on a superconducting substrate, Refs. ~\cite{Nyari23,Laszloffy23} provide an example of zero energy localized states which are not Majorana fermions.  An interesting case is that of superconductor semiconductor heterostructures in which odd-frequency superconductors~\cite{Tanaka12,Linder19,Cayao20} are used as a substrate.  The understanding of these systems requires the spectral BBC~\cite{Tamura19,Daido19,Tamura21,Ahmed25}.  A recent study~\cite{Ahmed24} identifies the signatures of Andreev bound states, also localized at the edges, and which are also either topological or trivial.  In OBC non-Hermitian systems edge localization occurs due to the non-Hermitian skin effect~\cite{Zhang22}, and possibly, due to other topological effects.  In all of these cases, the question becomes whether edge modes of topological origin can be distinguished from other types.  On the other hand, in interacting systems, edge states can smear out even in topologically nontrivial phases, manifesting as zeros of the Green's function~\cite{Gurarie11}.} \\ 

A major breakthrough on the BBC was the series of papers of Alase et al. \cite{Alase16,Alase17,Cobanera17,Cobanera18} who generalized the Bloch theorem (GBT) for systems in which translational invariance is broken by the presence of boundaries.  This approach first casts all terms in the Hamiltonian as a direct product of the lattice shift operator and "small" matrices which connect internal degrees of freedom (within a given unit cell, as well as between unit cells).   The Schr\"odinger equation is separated into bulk and boundary equations using projectors.  The bulk equation is simplified by diagonalizing the lattice shift operator, whose eigenvalues, $z$, all fall on the unit circle, for a periodic system.  The formalism also proposes a quantity, the BBC indicator ($D$), which is sensitive to localization (diverges), but is not directly sensitive to whether a state is topological or not.  We emphasize that this formalism is fundamentally the same as the non-Bloch band theory~\cite{Yao18,Yokomizo19} used in constructing the generalized Brillouin zone and the topological invariants of non-Hermitian systems~\cite{Bergholtz21}.   \\

Other, more direct methods have also been developed~\cite{Sticlet12,Sedlmayr15,Awoga24} to study edge states in topological systems.  For systems in which the edge mode is a Majorana fermion, the Majorana polarization~\cite{Sticlet12} is often the method of choice.  This method proposes a quantity in a restricted part of the real lattice, and calculates an expectation value of the particle-hole symmetry operator.  Recently this method has been demonstrated to work in higher dimensions~\cite{Sedlmayr15}.  In the presence of disorder evidence exits of its usefulness~\cite{Awoga24}, but it can not unambiguously determine~\cite{Karoliya25} whether a zero energy edge state is a Majorana fermion.\\

\textcolor{black}{In this work, we develop two extensions of the GBT formalism.  On the one hand, given that $D$ is only localization sensitive, and is not sensitive to topology, we use the symmetry protecting operator of topological edge states to filter the edge state bases of the GBT.  In addition, we calculate the open boundary Zak phase.  This development can be viewed as the continuation of two important and related previous ones: the generalized Wannier formalism and the GBT.  In both cases, a well-known concept used widely in PBC (crystalline) systems was extended crysalline OBC systems.  We do exactly this in the case of the Zak phase, which, until know, has only seen widespread use, but only in PBC systems.  We apply our developments to the Kitaev wire, mainly to demonstrate principle, and also to a system with a bond impurity, where we study the connection between localization and topology.}  \\

Our paper is organized as follows.  In the next section, we summarize the aims of our paper and state our results.  After that, intended as background, we present the derivation of the BBC indicator ($D$) of the GBT~\cite{Alase16,Alase17,Cobanera17,Cobanera18}.  In section \ref{sec:TI} we show how a Brillouin zone can be constructed for open systems and how a Zak phase (topological invariant) can be calculated.  In section \ref{sec:Kitaev} we present a calculation for a Kitaev wire.  In this section we also carry out a symmetry analysis.  In the penultimate section, we use a generalized version of both formailsms to study a periodic system with a bond impurity.  In the last section we conclude our work.\\

\section{Statement of our results}

\textcolor{black}{In this work we develop and test two extensions of the GBT~\cite{Alase16,Alase17,Cobanera17,Cobanera18}.   We complement the BBC indicator ($D$) of the GBT in two ways.   One, we apply the usual symmetry analysis that is ubiquitous in the analysis of topological insulators and superconductors~\cite{Altland97,Zirnbauer96,Schnyder08,Ryu10,Kitaev09}, by filtering the basis set for the edge states of the formalism using the relevant symmetry operator.  Our example calculation is for the Kitaev wire and the relevant symmetry operator, which protects the topoligcal edge states, is the chiral symmetry operator.  We calculate $D$ within such filtered basis sets.  Two, we develop a formalism for the actual topological invariant (the Zak phase~\cite{Berry84,Zak89}, $\gamma_Z$ in the case of the Kitaev model) for the OBC system.   The Zak phase was originally developed for systems with PBC, here we calculate a Zak phase formed along points in the Brillouin zone of the OBC system.   For our chosen example, the Kitaev wire, the different quantities we calculate show expected correspondence with known behavior: a divergent symmetry restricted $D$ in the known topologically non-trivial region is not only a sign of a localized edge state, but is also verified as topological, due to a non-trivial value of the Zak phase ($\gamma_Z = \pi$).   The converse holds in non-trivial regions (finite $D$ and $\gamma_Z=0$).  Our calculations also identify tendencies as a function of system size.  We also analyze the behavior of $z$ values, the eigenvalues of the lattice shift operator which contribute to the wavefunctions of a given system.  The behavior of the $z$ values associated with the edge states is also indicative of the transitions which occur, in a manner qualitatively similar to the behavior of Lee-Yang zeros~\cite{Lee52a,Lee52b,Timonin21,Chitov22} in classical phase transitions.  We comment on this question further.}\\

\textcolor{black}{In addition, motivated by the recent interest~\cite{Muller16,Tinyukova19,Mishra21,Woods21,Peeters24} in Kitaev wires with impurities, we also study a wire with periodic boundary conditions and with one bond impurity.   $D$ and the $\gamma_Z$ can be calculated with slight modifications of the original formalism.  For $D$, we find that a scan in the chemical potential ($\mu$) results in isolated point singularities at the topological phase transition and also at another value of $\mu$, where the entry into or exit out of the circle of oscillations~\cite{Hegde16,Alase17} region (COO) occurs.   The COO was first studied in detail by Hegde and Vishveshwara~\cite{Hegde16}.   Majorana edge states exhibit a gap induced decay, but they can also exhibit oscillations depending on the value of the chemical potential.   The region of $\mu$ identified as a topologically nontrivial region for the OBC system still exhibits $\gamma_Z = \pi$.  The interpretation here is that an impurity system is still topologically non-trivial, but is not accompanied by localization near the impurity to the same extent as the edge state of the OBC system.    Potentially, there is a theoretical advantage in numerical settings to using a bond impurity system to assess topologically non-trivial behavior, because isolated singularities are easier to identify than entire regions of $\mu$ where $D$ diverges. }

\section{Background: the bulk boundary separation method}

In this section we revisit the GBT formalism introduced by Alase et al. in Refs. \cite{Alase16,Alase17,Cobanera17,Cobanera18}.   \textcolor{black}{This section is intended as background and a starting point for our formal developments, which are presented in later sections.}   In particular, we derive the bulk-boundary correspondence indicator ($D$), the quantity sensitive to localization.  We wish to emphasize that this formalism is in a very close relation to the non-Bloch band theory~\cite{Yao18,Yokomizo19} used in the analysis of non-Hermitian systems.  In our description we will point out some of the essential similarities.\\

\begin{figure}[t]
 \centering
 \includegraphics[width=8cm]{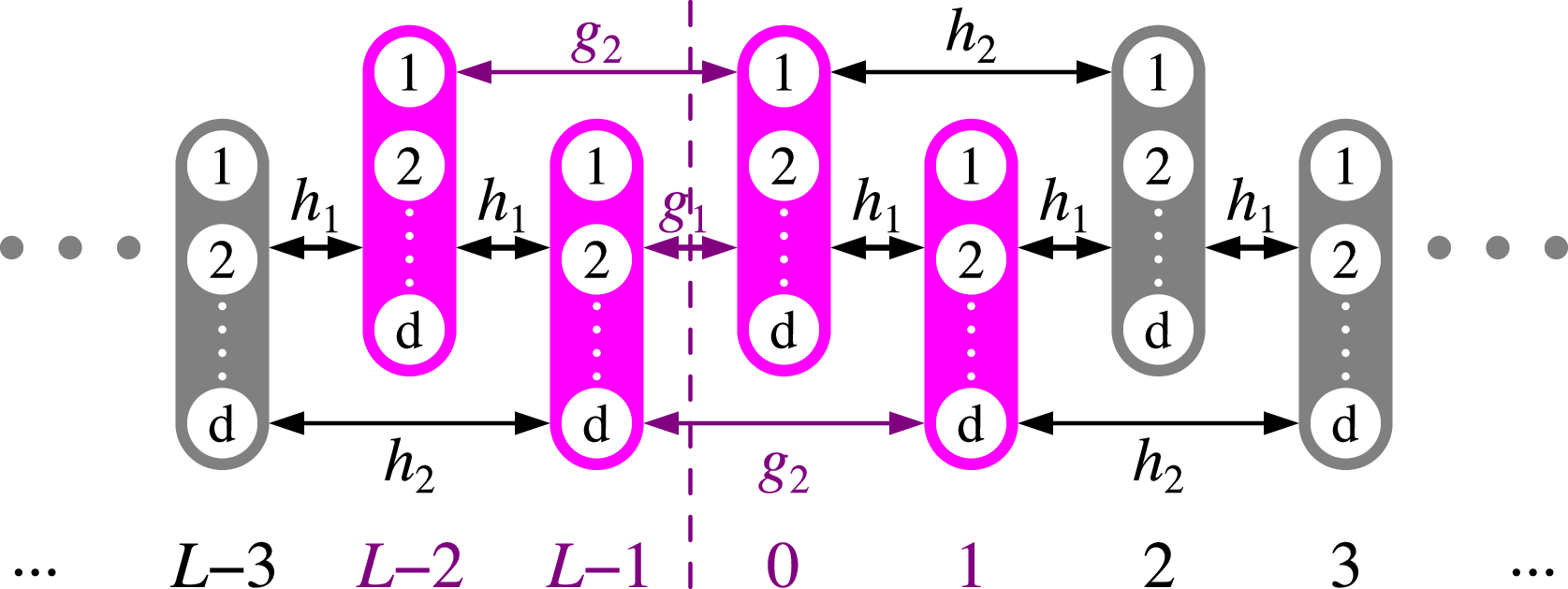}
 \caption{Schematic representation of a lattice corresponding to the grouping of the terms of the Hamiltonian in Eq. (\ref{eqn:H1}) for a case of $L$ lattice sites and range of $R=2$.   The figure represents a ring of sites indexed $0,...,L-1$.  The vertical sites respresent a lattice site with $d$ degrees of freedom.  When the Bogoliubov-de-Gennes representation is used, the number of degrees of freedom associated with a give lattice site doubles ($2d$).  The violet vertical stripes represent sites associated with the edges (lattice sites $L-2,L-1,0,1$).  The two-ended arrows represent couplings between lattice sites.  $h_1$ and $h_2$ denote couplings between bulk sites, while $g_1$ and $g_2$ represent couplings between boundary sites.  The boundary projector, $\hat{P}_\partial$, projects sites $L-2,L-1,0,1$ for a system of range $R=2$, while the bulk projector, $\hat{P}_B$ projects all other sites.  The case $g_1 = g_2 = 0$ corresponds to open boundary conditions.  $g_1 = h_1$ and $g_2 = h_2$ corresponds to periodic boundary conditions.  $g_1 \neq h_1$ and $g_2 \neq h_2$ corresponds to an periodic system with an impurity. }
 \label{fig:modelpic}
\end{figure}

Given a system of size $L$ with sites $j=0,...,L-1$ and internal degrees of freedom $m=1,...,d$.   We define the operator $\psi_j^\dagger = (a_{j1}^\dagger,...,a_{jd}^\dagger,a_{j1},...,a_{jd})$, where $a_{jm}$($a_{jm}^\dagger$) denotes an annihilation(creation) operator at site $j$ in internal state $m$.  We now write the Hamiltonian of this system as,
\begin{equation}
\label{eqn:H1}
\hat{H} = \frac{1}{2} \sum_{r=0}^R \left( \sum_{j=0}^{L-1-r} \psi_j^\dagger h_r \psi_{j+r} + \sum_{j=L-r}^{L-1} \psi_j^\dagger g_r \psi_{j+r-L} + \mbox{H.c.} \right),
\end{equation}
where the $2d \times 2d$ matrices $h_r$ and $g_r$ denote hopping and pairing matrices in the bulk and at the boundary, respectively.  \textcolor{black}{Most importantly, the matrices $h_r$ and $g_r$ connect internal degrees of freedom between different unit cells: $h_0$ and $g_0$ connect internal degrees of freedom on the same unit cell, $h_r$ and $g_r$ connect internal degrees of freedom between unit cells separated by $r$ units.  } $R$ denotes the range of the Hamiltonian, the largest distance between unit cells connected by hopping and/or pairing.   \textcolor{black}{This way of writing the Hamiltonian allows for a simple implementation of any boundary condition}.  Periodic boundary conditions correspond to $g_r = h_r$, while $g_r = 0$ for open boundaries.  The Hamiltonian, as written in Eq. (\ref{eqn:H1}), is represented in Fig. \ref{fig:modelpic}.  \\

It is possible to factorize the Hilbert space as $\mathcal{H} \simeq \mathbb{C}^L \otimes \mathbb{C}^{2d}$, a direct product for the lattice and the internal degrees of freedom.  We can now introduce the basis ${ | j \rangle | m \rangle }$, where $j= 0,...,L-1$ and $m = 1,...,2d$.  Basis states $| j \rangle | m \rangle$ with $m=1,..,d$ correspond to the creation operators $a^\dagger_{j,m}$, while the states with $m=d+1,...,2d$ to the annihilation operators, $a_{j,m}$.  The factorization of the Hilbert space allows the rewriting of the Hamiltonian in terms of the direct product of lattice shift operators, $\hat{T}$ (which acts between unit cells), and the "small" matrices $h_r$ and $g_r$ , as 
\begin{equation}
\label{eqn:H2}
\hat{H} = \sum_{r=0}^R [\hat{T}^r \otimes h_r + (\hat{T}^\dagger)^{L-r}\otimes g_r + \mbox{H. c.}],
\end{equation}
where the left-shift operator satisfies $\hat{T}|j\rangle = |j-1\rangle, \forall j \neq 0$, and $\hat{T}|0\rangle = 0$.  $\hat{T}^\dagger$ implements the corresponding right-shift.   \\

Periodic boundary conditions can be implemented by using the operators $\hat{V} = \hat{T} + (\hat{T}^\dagger)^{L-1}$.  The Hamiltonian now becomes
\begin{equation}
\hat{H} = \sum_{r=0}^R [\hat{V}^r \otimes h_r +  \mbox{H. c.}],
\end{equation}
and $\hat{H}, \hat{V}$ and $\hat{V}^\dagger$ from a commuting set.  Using the generalized $z$-transformed lattice basis, 
\begin{equation}
\label{eqn:z}
| z \rangle = \frac{1}{\sqrt{N(z)}} \sum_{j=0}^{L-1} z^j | j \rangle, z \in \mathbb{C}, z \neq 0.
\end{equation}
one can now obtain the "reduced bulk Hamiltonian",
\begin{equation}
\label{eqn:h_B}
h_B(z) = \sum_{r=0}^R (z^r h_r + z^{-r} h_r^\dagger).
\end{equation}
\textcolor{black}{In the most general case, $z$ is the parameter of the $z$ transform, but it is not the eigenvalue of the lattice shift operator, except} for a PBC system.   In this case, $z=e^{ik}$ is the eigenvalue of the shift operator ($\hat{V}$), and $z$s lie on the unit circle ($k$ is the crystal momentum).    The first Brillouin zone for a finite system with PBC is usually represented on the points $k = 2 \pi q/L$, $q = 0,...,L-1$.  The eigenvectors of $\hat{H}$, which are ordinary Bloch states, take the form $|\epsilon \rangle = |z \rangle |u(\epsilon, z)\rangle$, where $|u(\epsilon,z)\rangle$ is the eigenvector of $h_B(z)$ with eigenvalue $\epsilon$.   The $z$-transform in this case is a discrete Fourier transform.  \\

In the non-Bloch band theory of Refs. \cite{Yao18,Yokomizo19} the lattice shift operator, $\hat{T}$ is used implicitly.  In Ref. \cite{Yao18} the variables denoted as $\beta$ are eigenvalues of the translation operator, the analogs of $z$.   In our case, for PBC, $|z|=1$, for OBC, $|z|=1$ or $\mbox{Im}z= 0$, the only exception being  the topological edge states (discussed below).  In a non-Hermitian system the possible values of $z$ are more diverse~\cite{Yao18,Yokomizo19}, but the generalized Brillouin zone encircles the origin, albeit, it is not a unit circle.\\

For systems with open boundary conditions, where lattice translation symmetry is broken at the boundaries, the Fourier transform fails to diagonalize $\hat{H}$, and the operators $\hat{T}$ and $\hat{T}^\dagger$ do not share a common eigenbasis.  In the GBT the range of $z$ is first extended to the entire complex plane, allowing for decaying edge states at the boundaries.  Second, the eigenvalue equation is split into two pieces, a bulk and a boundary equation, using two projector operators, the bulk ($\hat{P}_B$) and the boundary ($\hat{P}_\partial$) projectors, defined as
\begin{eqnarray}
\hat{P}_B &=& \sum_{j=R}^{L-1-R} | j \rangle \langle j |, \\
\hat{P}_\partial &=& \mathbb{I}_L - \hat{P}_B, \nonumber
\end{eqnarray}
where $\mathbb{I}_L$ denotes the identity in the Hilbert subspace associated with lattice sites.   Correspondingly, the bulk and boundary equations read, 
\begin{eqnarray}
\hat{P}_B \hat{H} | \epsilon \rangle &=& \epsilon \hat{P}_B |\epsilon \rangle, \\
\hat{P}_\partial \hat{H} | \epsilon \rangle &=& \epsilon \hat{P}_\partial |\epsilon \rangle, \nonumber
\end{eqnarray}
respectively.  As a result of this separation, we obtain simultaneous relative eigenvectors of the bulk-projected $\hat{T}$ and $\hat{T}^\dagger$ operators.  The "generalized Bloch states" which will serve as the basis for the diagonalization procedure are of the same product form, $|z\rangle |u(\epsilon,z)\rangle$, where, again, $|u(\epsilon,z)\rangle$ is the eigenvector of $h_B(z)$ with eigenvalue $\epsilon$, but now $z$ is extended to the entire complex plane. \\

\subsection{Bulk-boundary correspondence indicator}

Putting the above into practice, one first solves the reduced bulk equation, which is a characteristic polynomial equation of $h_B(z)$,
\begin{equation}
\label{eqn:characteristic}
P(\epsilon,z) = z^{2dR} \mbox{det}|h_B(z) - \epsilon \mathbb{I}_{2d}| = 0,
\end{equation}
where $\mathbb{I}_{2d}$ denotes the identity in the $2d$-dimensional Hilbert space of the internal degrees of freedom, and the prefactor $z^{2dR}$ ensures that $P(\epsilon,z)$ is a bivariate polynomial in $\epsilon$ and $z$ of degree at most $4dR$.  Let us assume that there are $n$ values of $z_l$ which satisfy Eq. (\ref{eqn:characteristic}) for a given $\epsilon$ ($l=1,...,n$), and that for a given $z_l$, there are $s_l$ linearly independent eigenvectors of $h_B(z_l)$.  The full solution for a given eigenvalue $\epsilon$ can be written,
\begin{equation}
|\epsilon \rangle = \sum_{l=1}^n \sum_{s=1}^{s_l} \alpha_{l,s} | z_l(\epsilon) \rangle |u_s(\epsilon,z_l) \rangle, \alpha_{l,s} \in  \mathbb{C}.
\end{equation}
The coefficients $\alpha_{l,s}$ are determined from solving the boundary equation, $\hat{P}_\partial (H - \epsilon \mathbb{I})|\epsilon \rangle = 0$.   One can derive an equation for $\alpha_{l,s}$ by first setting up a basis for the $4dR$ boundary states $\{|j\rangle|m\rangle, 0\leq j \leq R-1, L-R \leq j \leq L-1 ; 1 \leq m \leq 2d \}$.  We can then form the $4dR$ equations,
\begin{equation}
\langle j | \langle m | \hat{P}_\partial (\hat{H} - \epsilon \mathbb{I})  | z_l(\epsilon) \rangle |u_s(\epsilon,z_l) \rangle \alpha_{l,s} = 0.
\end{equation}
For the edge states of topological origin, $\epsilon$ is usually zero.  One can form a matrix,
\begin{equation}
[B_L]_{jm;ls} = \langle j | \langle m | \hat{P}_\partial H  | z_l(0) \rangle |u_s(0,z_l) \rangle.
\end{equation}
The bulk-boundary indicator defined in Ref. \cite{Alase16} reads as,
\begin{equation}
\label{eqn:D}
D = \log \det \{ B^\dagger_\infty B_\infty \}.
\end{equation}
In sections \ref{sec:Kitaev} and \ref{sec:KitaevBondImp} for the Kitaev wire type models, we will calculate a modfied version of Eq. (\ref{eqn:D}).  Each member of the edge state basis, $ | z_l(0) \rangle |u_s(0,z_l) \rangle $ is an eigenstate of the chiral symmetry operator, the operator which, in the Kitaev wire, protects the topological edge states.  We will calculate $D$ in a basis in which the basis states all have the same chiral symmetry eigenvalue.\\

An OBC or PBC system are special cases of any lattice model of the tight-binding type with a bond impurity (see Eq. (\ref{eqn:tDl})).  This means that, with minor modifications, the formalism derived above can be applied to a wire with a bond impurity.  This is done in section \ref{sec:KitaevBondImp}. \\

\subsection{Connection to Lee-Yang zeros}

\textcolor{black}{Below, we will also make an analysis of $z$ values, which, as we remarked earlier, is similar to the analysis of Lee-Yang zeros\cite{Lee52a,Lee52b,Timonin21,Chitov22} in classical phase transitions.   While the relation of the Lee-Yang approach to the GBT formalism is an interesting question, it deserves further study and is beyond the scope of this work.  However, a few basic remarks are in order.  In the Lee-Yang approach one applies a complex magnetic field in a classical Ising system, which modifies the Boltzmann factor by  $\tilde{z}= \exp(\beta h \sigma)$, where $\beta$ denotes the inverse temperature, $h$ denotes the complex magnetic field, and $\sigma$ denotes the spin variables.  Using this modification of the Boltzmann factor, one searches for the roots of the partition function as a function of $\tilde{z}$.  For comparison, note that the $z$-transform applied above to a quantum lattice Hamiltonian has the form $z=\exp(i \kappa j)$, where $\kappa$ is a momentum, in general complex, and one solves for the roots of a secular equation.  While there are other differences, we remark than one important difference is that $z$ can be interpreted as the application of a complex electric field, rather than a complex magnetic field, since the position (lattice site $j$) appears explicitly in $z$ but not in $\tilde{z}$. } \\

\section{Calculating the topological invariant for a system with open boundaries}

\label{sec:TI}

\textcolor{black}{The question of relating OBC and PBC systems of a given model is not new.  For continuous models generalized Wannier functions for OBC systems have been introduced~\cite{Kohn73,Rehr74,Kivelson82,Nenciu98} which are localized and asymptotically approach the usual (PBC) Wannier functions.  The GBT was developed for lattice systems, and is based on the generalization of the Bloch theorem (a theorem originally proposed for PBC systems) for OBC systems.  The Zak phase~\cite{Zak89} is also a quantity that was derived for PBC systems.  In the spirit of the two developments mentioned, we aim to generalize the Zak phase to OBC systems.  An intercellular Zak phase~\cite{Rhim17} for a PBC system, from which the charges on the edge of an OBC system can be determined, has also been derived.}

In one dimensional (1D) topological condensed matter systems~\cite{Bernevig13,Asboth16,Kane05a,Kane05b,Kitaev01}, topological invariants are Zak phases~\cite{Zak89} or modifications thereof.  In the Su-Schrieffer-Heeger model~\cite{Su79}, perhaps the first and simplest topologically non-trivial model, the topological invariant is an ordinary Zak phase which in turn also corresponds to the bulk polarization~\cite{King-Smith93,Resta94,Resta00,Vanderbilt18}.  The canonical spinful time-reversal invariant 1D topological insulator is the Fu-Kane $\mathbb{Z}_2$ topological insulator~\cite{Fu06}, whose topological invariant is known as time reversal polarization.  The time reversal polarization can be expressed as the difference of two Zak phases in the different spin channels, but also, as the product of two Pfaffians at the time reversal invariant $k$-points, as shown in Refs \cite{Bernevig13,Fu06}.   The Majorana number, the topological invariant for the Kitaev wire, derived by Kitaev, is also a product of Pfaffians, however, the Zak phase picked up by integrating over the Brillouin zone is equivalent to the Pfaffian based expression~\cite{Budich13}.  When the Kitaev 1D $p$-wave SC model is cast in terms of Bloch states (Eq. (13) of Ref. \cite{Kitaev01}) the result is identical to the SSH model.   The Zak phase is still a valid topological invariant, but, unlike in the SSH model, it does not correspond to the bulk polarization.  \\

 In this section we describe how one can calculate the Zak phase for an OBC system, generalizing the original~\cite{Zak89} PBC system.   One way to write the Zak phase in a PBC system is to use the periodic Bloch wavefunctions, $|u_k\rangle$, as a Bargmann invariant~\cite{Bargmann64}, which for a system of size $L$ can be written as,
 \begin{equation}
 \gamma_Z = \prod_{m=1}^L \langle u_{k_m} | u_{k_{m+1}} \rangle,
 \end{equation}
 where $k_m$ denote a set of $k$-points which are evenly spaced within the Brillouin zone as $k_m = 2 \pi m /L$.  We will construct a Bargmann invariant in the effective Brillouin zone of the OBC system.  The $k$- values will not be necessarily evenly spaced, but at each $k$ value (or $z$ value, with $|z|=1$) we will construct the appropriate $z$-dependent "small" wavefunction. \\

Consider a Hamiltonian $\hat{H}_{\mbox{o}}$ with open boundary conditions, arranged according to the form of Eq.  (\ref{eqn:H2}) (with $g_r=0$).   Importantly, such a Hamiltonian also has an associated reduced Hamiltonian, $h_B(z)$, according to Eq. (\ref{eqn:h_B}).  We also assume that  $\hat{H}_{\mbox{o}}$ has an eigensystem,
\begin{equation}
\label{eqn:Ho}
\hat{H}_{\mbox{o}} |\epsilon_J \rangle = \epsilon_J |\epsilon_J \rangle, J = 1,...,2Ld
\end{equation}
where $L$ is the number of unit cells, $d$ is the number of internal degrees of freedom.  We also assume that the eigensystem can be separated into bands.  For example, let the lowest $J=1,...,N$ from a band.  In principle, midgap states, such as edge states (which are usually degenerate), can affect this separation.  However, since the Zak phase is formed from states with $|z|=1$, such states can be excluded. \\

The topological invariant is a Zak phase, an integral around the Brillouin zone.  Accordingly, what is needed is to find the components of the wavefunctions of the given band for which $z$ lies on the unit circle.  For each  $\epsilon_J$, $J=1,...,N$, we solve the reduced bulk Hamiltonian equation,
\begin{equation}
\label{eqn:characteristic_jl}
P(\epsilon,z_{J\lambda}) = z^{2dR} \mbox{det}|h_B(z_{J\lambda}) - \epsilon_J \mathbb{I}_{2d}| = 0.
\end{equation}
For each $\epsilon_J$ we obtain $z_{J\lambda}$, $\lambda=1,...,4dR$ (they are distinct, unless the system is at a phase transition point, where the Zak phase is undefined).  \\

The next step is to keep only those $z_{J\lambda}$, for which $|z_{J\lambda}|=1$.  Suppose that we find $N_k$ such values of $k_M$ ($M = 1,...,N_k$).  In addition, we put these in increasing order according to the crystal momentum $k_M$, and we also keep track of the original energy eigenstates which correspond to a given $k_M$, so we have  $(k_M,\epsilon_{J_M}), M = 1,...,N_k$.  Note that more than one $k_M$ can correspond to the same energy eigenvalue.   At each $k_M$ we form $2d$ wave functions of the form,
\begin{equation}
\label{eqn:A}
|A_\alpha(k_M) \rangle = |z=e^{ik_M}\rangle  |\hat{e}_\alpha \rangle,
\end{equation}
where $|\hat{e}_\alpha\rangle$ denotes a basis vector (column) of unit length in the $2d$ dimensional space of the internal degrees of freedom in the direction $\alpha$.  We now form a $2d$ dimensional vector for each $k_M$,
\begin{equation}
\label{eqn:psikm}
|\Psi(k_M)\rangle = \begin{pmatrix} \langle A_1(k_M) | \epsilon_{J_M} \rangle \\.\\.\\.\\\langle A_{2d}(k_M)|\epsilon_{J_M} \rangle \end{pmatrix}
\end{equation}
Using the $2d$-dimensional vectors $|\Psi(k_M)\rangle$, we form the discrete Zak phase (Bargmann invariant~\cite{Bargmann64}), by way of the cyclic product,
\begin{equation}
\gamma_Z = \mbox{Im} \ln \prod_{M=1}^{N_k} \langle \Psi(k_M)| \Psi(k_{M+1}) \rangle,
\end{equation}
where $k_{N_k+1}=k_1$.  We also check whether the weight of a given function at a particular $k_M$, $\langle \Psi(k_M)| \Psi(k_M) \rangle$, is greater than zero, otherwise they would not be included in the calculation. \\

Our construction makes two points of contact with the eigensystem obtained from the original, open boundary, Hamiltonian.  One is that the $z$ values are obtained from substituting $\epsilon_J$ for the band of interest into the secular equation of the "small Hamiltonian", Eq. (\ref{eqn:characteristic}).  The other is that $|\Psi(k_M)\rangle$ is constructed by taking overlaps between the states $|\epsilon_J\rangle$ with states of the $z$-values obtained from solving  Eq. (\ref{eqn:characteristic}).  This last part amounts to extracting the components of $|\epsilon_J \rangle$ whose $z$-values correspond to the Brillouin zone.  The formalism derived here can be applied without modification to the Kitaev wire with PBC and a bond impurity. \\

\section{Kitaev wire with open boundary conditions}

\label{sec:Kitaev}

Our first example is the spinless Kitaev wire, whose Hamiltonian in the OBC case is given by,
\begin{equation}
H_K = \sum_{j=0}^{L-1} \left( -w a_j^\dagger a_{j+1} - \mu a_j^\dagger a_j + \Delta a_j^\dagger a_{j+1}^\dagger + \mbox{H. c.} \right).
\end{equation}
where $w$ denotes the hopping, $\Delta$ denotes $p$-wave pairing, and $\mu$ denotes the chemical potential.  The operators $a_j$($a_j^\dagger$) denote the annihilation(creation) operators at site $j$, and $L$ denotes the system size.  It is known that the topological phase occurs for $|\mu| < 2 |w|$.  Also, pertinent to our study is that $H_K$ exhibits a COO defined as 
\begin{equation}
\left( \frac{\mu}{2w} \right)^2 + \left( \frac{\Delta}{w} \right)^2 = 1.
\end{equation}
Within this circle the Majorana modes decay and oscillate (similar to underdamped oscillators), while outside this circle, but still within the topologically nontrivial phase, the Majorana modes only decay. \\

For this model $d=R=1$, and the small matrices for the form of the Hamiltonian of the form in Eq. (\ref{eqn:H2}) can be written as,
\begin{equation}
h_0 = \begin{pmatrix} -\mu & 0 \\ 0 & \mu \end{pmatrix}, h_1 = \begin{pmatrix} -w & \Delta \\ -\Delta & w \end{pmatrix}.
\end{equation}
\textcolor{black}{The reduced bulk Hamiltonian (of the form in Eq. (\ref{eqn:h_B})) reads,}
\begin{equation}
h_B(z) = \begin{pmatrix} -\mu - w (z + z^{-1})& \Delta(z - z^{-1}) \\ -\Delta(z-z^{-1}) & \mu + w(z + z^{-1}) \end{pmatrix}.
\end{equation}
\\

\textcolor{black}{For the symmetry analysis we need to write the relevant symmetry operators, which are time reversal ($\hat{\mathcal{T}}$), particle-hole ($\hat{\mathcal{P}}$) and chiral ($\hat{\Sigma}$), in the same form as Eq. (\ref{eqn:H2}).  These are,}
\begin{equation}
\hat{\mathcal{T}} = K, \hspace{.5cm} \hat{\mathcal{P}} = (\mathbb{I}_L \otimes \sigma_x) K, \hspace{.5cm} \hat{\Sigma} = \mathbb{I}_L \otimes \sigma_x,
\end{equation}
\textcolor{black}{respectively, where $K$ denotes the complex conjugation operator, $\mathbb{I}_L$ denotes the identity operator on the lattice, and 
\[
\sigma_x = \begin{pmatrix} 0 & 1 \\ 1 & 0 \end{pmatrix}.
\]   
The model falls in the BDI symmetry class, since $\hat{\mathcal{T}}^2=1$, $\hat{\mathcal{P}}^2 = 1$.    Since the chiral symmetry operator anti-commutes with the Hamiltonian, it is guaranteed that for every eigenstate with energy $\epsilon$, there exists one with energy $-\epsilon$.  A further consequence of anti-commutation is that the zero energy edge states are also eigenstates of the chiral symmetry operator (for a proof, see Ref. \cite{Gurarie11}).}

\begin{figure}[t]
 \centering
 \includegraphics[width=7cm,keepaspectratio=true]{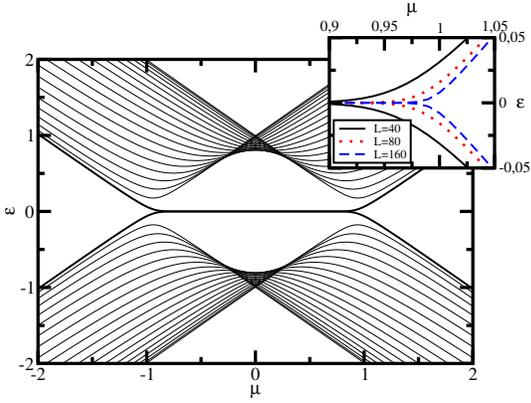}
 \caption{Kitaev wire with open boundary conditions.  Band structure (energies , $\epsilon$ as a function of the chemical potential, $\mu$) for a system with size is $L=20$.   A zero energy edge state forms at  a chemical potential of $\mu\approx -1$ and persists until $\mu \approx 1$.  Due to the finite (and small) system size the edge states are not exactly in the range $|\mu| \leq 1$, as it occurs in the thermodynamic limit.  The inset shows the opening of the gap as a function of $\mu$ for three different system sizes, $L=40, 80$ and $160$.  Clearly, as the system size increases, the point to which the degeneracy persists approaches $\mu=1$.}
 \label{fig:bsE0}
\end{figure}

\textcolor{black}{One can also apply the $z$-transform to the symmetry operators, in which case, the "reduced bulk" version of the chiral symmetry operator ($\hat{\sigma}_B$) becomes,}
\begin{equation}
\hat{\sigma}_B = \sigma_x,
\end{equation}
\textcolor{black}{which anticommutes with $h_B(z)$ for any value of $z$.   It is fairly simple to show that a zero energy eigenvalue of $h_B(z)$ is an eigenstate of $\hat{\sigma}_B$, therefore, the basis states for edge state, $| z \rangle |u(z) \rangle$ are also eigenstates of the chiral symmetry operator. }\\
\begin{figure}[t]
 \centering
 \includegraphics[width=8cm]{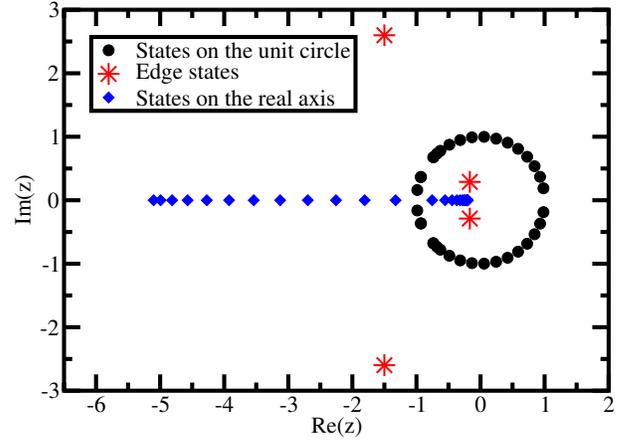}
 \caption{Kitaev wire with open boundary conditions.  $z$ values for a system with $L=16$ and (hopping $t=0.5$, superconducting pairing $\Delta=0.4$, and chemical potential $\mu=0.3$) a system which is topological and is within the COO.  The black dots indicate states for which $|z|=1$.  We use these states (black dots) to calculate the topological invariant $\gamma_Z$.  The blue diamonds are states with $\mbox{Im}(z)=0$.  The zero energy edge states (Majorana modes) are indicated with red asterisks.  For these states the momentum $k$ ($z=e^{ik}$) is complex as long as the system is within the COO.}
 \label{fig:states}
\end{figure}

The parameters for our model calculations are $w=0.5$, $\Delta=0.4$, and we will scan in $\mu$.  The known phase transition point in this case is $|\mu|=1$.   For these parameters the region $|\mu|<0.6$ is within the COO.\\

In Fig. \ref{fig:bsE0} we show the energy levels of a system with $L=20$ as a function of $\mu$.  The well known phase transition points occur near $\mu = 1$ and $\mu=-1$, and the figure clearly shows the zero energy states between those values, albeit, they do not form exactly at the known transition points ($|\mu|=1$), because $L=20$ is a finite system.  The inset shows the opening of the gap at zero energy as a function of  $\mu$ for different system sizes.   It is seen that the transition point is converging to $\mu=1$ as the system size is increased.  We want to further investigate this size dependence by other means. \\

In Fig. \ref{fig:states} the $z$ values \textcolor{black}{(parameters of the $z$ transform)} are shown on the complex plane for a system of size $L=16$ and $\mu=0.3$.   The $z$ values are found by solving Eq. (\ref{eqn:characteristic}) for each value of $\epsilon_J \leq 0$.   For each single eigenvalue, $\epsilon_J$, four $z$ values are found.  We find three different categories of $z$ values, which we indicate in the figure using different symbols and colors.  The $z$ values which comprise the Brillouin zone ($|z|=1$) are shown as black filled circles.  The topological invariant ($\gamma_Z$) in our scheme is calculated using such $z$ values, by extracting from each wavefunction the corresponding basis components, \textcolor{black}{(see Eq. (\ref{eqn:psikm}))}.   \textcolor{black}{A PBC system would only exhibit states for which $|z|=1$.}   \textcolor{black}{However, for our OBC system,} we also find a set of states for which $z$ only has a real part.  These are exponentially decaying and growing states (they come in reciprocal pairs) \textcolor{black}{which allow for localization at the edges.}  In addition, there are four $z$ values which correspond to the topological zero energy edge states (Majorana zero modes).   These are indicated with red asterisks.  These states have both a real and an imaginary part and they only occur for a system that is within the  of COO.  \textcolor{black}{These are the zero energy edge states of topological origin.}\\
\begin{figure}[t]
 \centering
 \includegraphics[width=8cm,keepaspectratio=true]{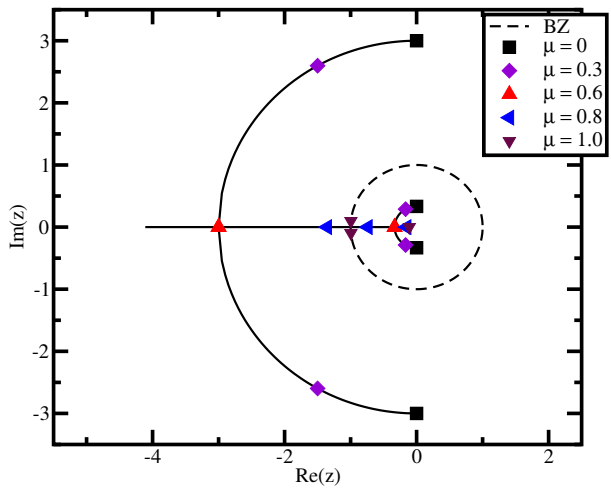}
 \caption{Kitave wire with open boundary conditions.  Solid lines represent the $z$ values for the state that is a zero energy edge state (in the topological phase) obtained from a scan in the chemical potential $\mu$ from $\mu=0$ to $\mu=1.2$.  The dashed lines represent the circle, $|z|=1$ which is the Brillouin zone.  To illustrate the direction of the flow of the lines, we show the $z$ values for edge states at particular values of $\mu$.}
 \label{fig:paths}
\end{figure}

We also investigated how the edge states evolve as the chemical potential $\mu$ changes.   These results are presented in Fig. \ref{fig:paths}.  The black lines are the result of a scan from $\mu=0$ to $\mu=1.2$.  We also show states for specific values of $\mu$ (solid black squares ($\mu=0$), solid violet diamonds ($\mu=0.3$),  solid red upward triangles ($\mu = 0.6$), solid blue triangles left ($\mu=0.8$), and solid brown triangles down ($\mu=1.2$)).  As $\mu$ is increased \textcolor{black}{from $\mu=0$ to $\mu=0.6$ (within the COO)}, the $z$ values of the Majorana zero modes maintain their magnitude, but all move simulatenously along a circle towards the real axis.  Once the boundary of the COO is reached, the $z$ values form two pairs of degeneracies on the real axis (two solid upward triangles at $(-3,0)$ and two at $(-1/3,0)$).  Increasing $\mu$ further splits these $z$-degeneracies again with the formerly degenerate $z$-values moving away from each other, but this time along the real axis.  The higher of the $(-3,0)$ pair and the lower of the $(-1/3,0)$ pair move towards each other, until the topological phase transition occurs at $\mu=1$, where these two $z$ values coalesce to form a degeneracy at $z=-1$ (which is on the Brillouin zone).  After this point, these states are no longer edge states, they have now acquired Bloch state characteristics.   Increasing $\mu$ further splits the degeneracy again, the two $z$ values move away from each other, but along the unit circle.\\ 

\begin{figure}[t]
 \centering
 \includegraphics[width=8cm,keepaspectratio=true]{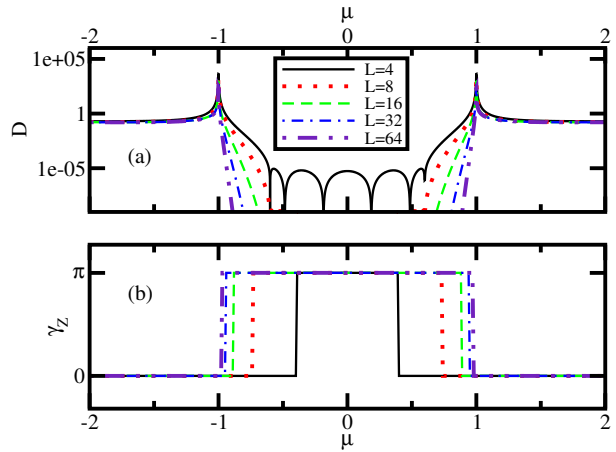}
 \caption{Kitaev wire with open boundary conditions.  Upper panel (panel (a)) shows the symmetry resolved bulk-boundary indicator, $D$, as a function of the chemical potential $\mu$ for system sizes $L=4,8,16,32,64$.  The vertical axis is logarithmic.  Lower panel (panel (b)) shows the topological invariant (Zak phase, $\gamma_Z$) for the same system sizes.}
 \label{fig:DvsZ}
\end{figure}

Fig. \ref{fig:DvsZ}(a) shows the bulk-boundary correspondence indicator, $D$ (Eq. (\ref{eqn:D})).  The $y$ axis in the figure is logarithmic.   Different system sizes are shown, starting from the very small size of $L=4$.  As a function of system size convergence can be seen towards the known transition points.  Results for the topological invariant ($\gamma_Z$) are also shown for the same system sizes (Fig. \ref{fig:DvsZ}(b)).  In the thermodynamic limit, the topological invariant takes the value $0$ in the trivial and $\pi$ in the non-trivial phase.  Again, this limit is clearly approached in the lower panel of Fig. \ref{fig:DvsZ}(b), however, the regions for which the topological invariant is nontrivial are smaller in this case than for the quantity $D$ (Fig. \ref{fig:DvsZ}(a)).   \textcolor{black}{The jump in $\gamma_Z$ corresponds entirely to the "jumping" of the $z$ values associated with the edge states "onto" the Brillouin zone.}  For the system with $L=4$, $\gamma_Z$ shows topologically nontrivial behavior only for $|\mu|<0.4$, whereas when $D$ is used as a criterion this region appears to be wider, approximately, $|\mu|<0.6$.  For other sizes, $\gamma_Z$ shows broader regions in which the behavior of the system is topologically nontrivial.   Both $D$ and $\gamma_Z$ indicate the known transition value of $|\mu|=1$ as the system size is increased.\\

As a final test we apply our formalism to a periodic system.  This leads to the usual topological invariant.  In this case there is no size dependence, even for $L=4$ the topological invariant is $\pi$ for $|\mu|<1$, and zero in the trivial region.  These results are not shown.\\

\section{Periodic Kitaev chain with bond impurity}

\label{sec:KitaevBondImp}

\textcolor{black}{Historically, the generalized Wannier formalism was first developed~\cite{Kohn73} for a system with one impurity, and then~\cite{Rehr74} to an OBC system.  The GBT was first developed for an OBC system~\cite{Alase16,Alase17,Cobanera17}, but it was soon applied~\cite{Cobanera18} to an impurity system.  In this section, we test our developments for a model with an impurity: a Kitaev wire with PBC in which one bond has altered parameters.  The parameter which characterizes the extent to which that one bond is altered is $\lambda$ and the parameters on the impurity bond are defined as:}
\begin{eqnarray}
\label{eqn:tDl}
w &\rightarrow& w(1-2\lambda) \\
\Delta &\rightarrow& \Delta(1-2\lambda). \nonumber
\end{eqnarray}
$\lambda=0$($\lambda=0.5$) corresponds to PBC(OBC).

\begin{figure}[t]
 \centering
 \includegraphics[width=8cm,keepaspectratio=true]{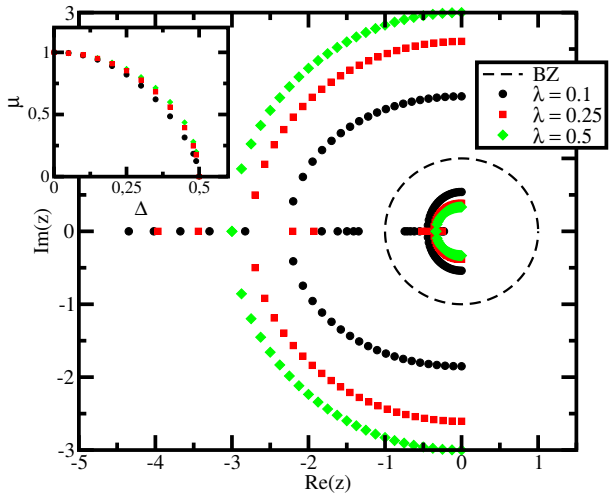}
 \caption{Kitaev model with one bond impurity.  Main panel shows the $z$ values of edge states resulting from a scan of $\mu$ (from zero to $0.6$) for different values of $\lambda$.   The system is of size $L=64$.  The values of $t$ and $\Delta$ for one bond in a ring is tuned according to Eqs. (\ref{eqn:tDl}).   $\lambda=0.5$ corresponds to open boundary conditions.   Inset shows the circle of oscillations.}
 \label{fig:pathsL}
\end{figure}

In Fig. \ref{fig:pathsL} the evolution of the $z$ values of the state which would be the highest occupied state at half filling (and also the Majorana edge state for a system with OBC) is shown for three cases $\lambda = 0.1, 0.25, 0.5$, as the variable $\mu$ is scanned from zero to  $0.6$ in steps of $0.025$.  Each such set of data (each value of $\lambda$) forms two quasicircular shapes, one inside, and one outside the Brillouin zone ($|z|=1$)).  For open boundary conditions the shapes are circles with radii $3$ and $1/3$, while for decreasing $\lambda<0.5$, the circles become elliptical, and while also being closer to the circle represented by the Brillouin zone.  When $\lambda=0$ (not shown), the states become part of the Brillouin zone $|z|=1$, because periodic boundary conditions are restored.  The modified Majorana states at $\lambda<0.5$ also exhibit a COO, a $\mu$ value above which the states lose their oscillatory character.  The inset of Fig. \ref{fig:pathsL} shows how the COO is modified for different values of $\lambda$.  \\

\begin{figure}[t]
 \centering
 \includegraphics[width=8cm,keepaspectratio=true]{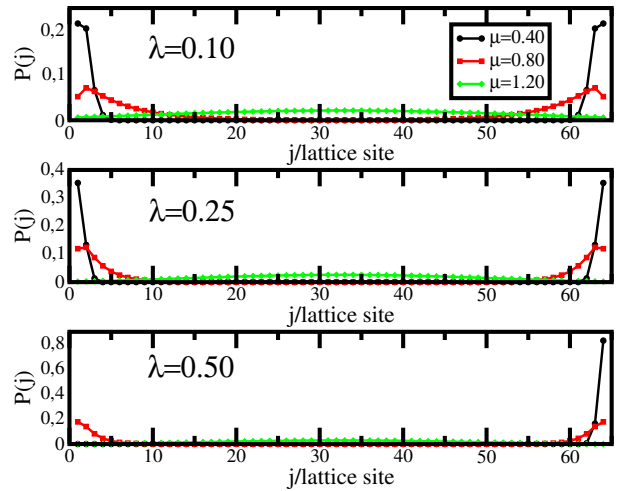}
 \caption{Evolution of the edge state as a function of $\mu$ and $\lambda$ in a Kitaev wire with one bond impurity.  For any value of $\lambda$, the states are localized at the edge within the topological phase ($\mu=0.40, 0.80$, albeit less so, if the system is outside the circle of oscillation ($\mu=0.80$).  The states are delocalized over the lattice for $\mu=1.20$.  $\lambda=0.5$ corresponds to open boundary conditions, here the topological edge state within the circle of oscillation is localized on one of the edges.}
 \label{fig:tD1ImpEdge}
\end{figure}

In Fig. \ref{fig:tD1ImpEdge} we show the probability distribution in the particle channel.   Each panel corresponds to a given value of $\lambda$, and results are shown for $\mu=0.4,0.8,1.2$.  The first value ($\mu=0.4$) is within the topological phase, as well as within the COO, the second ($\mu=0.8$) is within the former but outside the latter, while $\mu=1.2$ is outside of both.  For $\lambda=0.5$ the topological edge state is well localized at one edge, if the system is within the COO, while for $\mu=0.8$ we see localization at both edges.  Outside the topological region ($\mu=1.2$) the wave function is delocalized.  For $\lambda \neq 0.5$, for $\mu<0.6$ the state shows localization, but no longer only on one side.  Other than this, the localization tendencies are similar to the $\lambda=0.5$ case (for $\mu = 0.8$ and $\mu = 1.2$).  \\

\begin{figure}[t]
 \centering
 \includegraphics[width=8cm,keepaspectratio=true]{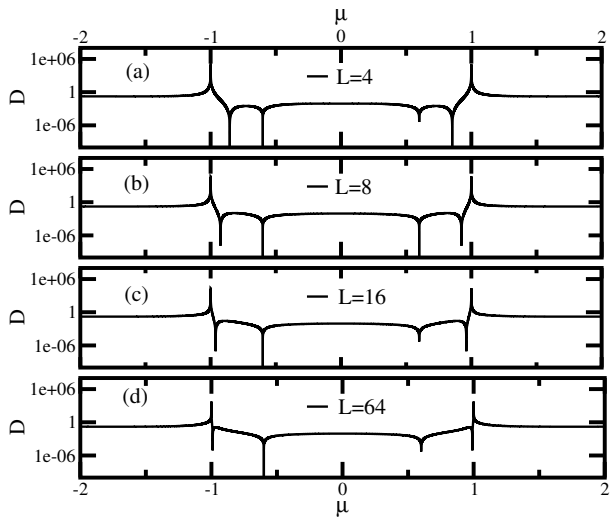}
 \caption{Bulk-boundary correspondence indicator ($D$, symmetry resolved) for four different system sizes as a function of the chemical potential $\mu$ for a Kitaev wire with periodic boundary conditions and one bond impurity ($\lambda=0.25$).   The $y$-axis is logarithmic.  The $x$-axis for all four plots shows the $\mu$.  All four system sizes are finite and continuous, except for six singular "spikes", three on the negative, three on the positive side.  Two of the spikes are "upwards", while four are "downwards".}
 \label{fig:bondimpDvsZ}
\end{figure}

The bulk-boundary correspondence indicator, $D$, \textcolor{black}{symmetry resolved (the eigenvalue of the chiral symmetry operator, $\hat{\Sigma}$, is unity)} is shown in Fig. \ref{fig:bondimpDvsZ} for four different system sizes as a function of $\mu$ for the case $\lambda=0.25$.  The scan is from $\mu=-2$ to $\mu=2$.  $D$ is symmetric with respect to $\mu \rightarrow -\mu$.  The important information of this figure is the spikes in $D$, of which, there are six, three for $\mu<0$, three for $\mu>0$.  All system sizes show two upward spikes (at $\mu = -1$ and $\mu = 1$), irrespective of system size.  Of the remaining downward spikes, two are at $\mu = 0.6$ and $\mu = -0.6$, also independent of system size.  The remaining downward spikes are near $\mu=-1$ and $\mu=1$.  The positions of these spikes show size dependence, and they approach $\mu = |1|$ as the system size increases.  This means that $D$ is singular at $\mu = |1|$ (at the topological phase transition) as well as at the transition when the system enters or exits the COO ($\mu = |0.6|$), but the nature of the singularities differs between the two types of transitions.  The former is similar to the singularity of the function $1/x$ at $x=0$, since different limits are approached depending on whether zero is approached from the negative or the positive side, while the latter is similar to the singularity of the function $-1/x^2$ in which case minuss infinity is approached irrespective of whether $x \rightarrow 0^+$ or $x \rightarrow 0^-$.\\

\begin{figure}[b]
 \centering
 \includegraphics[width=9cm,keepaspectratio=true]{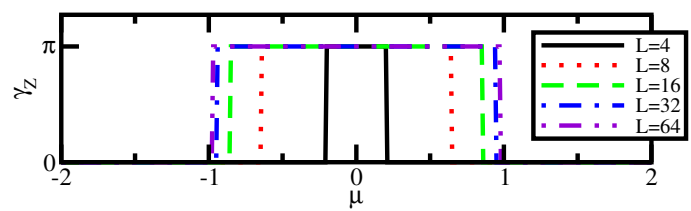}
 \caption{Kitaev wire with a bond impurity.  Open boundary Zak phase ($\gamma_Z$) as a function of the chemical potential, $\mu$, for different system sizes.}
 \label{fig:bondimpZ}
\end{figure}

Fig. \ref{fig:bondimpZ} shows the OBC Zak phase ($\gamma_Z$) as a function of the chemical potential, $\mu$ for the Kitaev wire with a bond impurity ($\lambda=0.25$).  For small system sizes, the region in which $\gamma_Z=\pi$ shrinks, but as the system size increases, the transition region increases to its PBC size ($[-\mu,\mu]$). \\

\textcolor{black}{It is in order to summarize the differences between the OBC system, studied in the previous section, and the PBC system with one bond impurity.   Overall, the bond impurity reduces the localization of the edge state, which can be seen in the behavior of the bulk-boundary correspondence indicator $D$, which is singular in the entire topological phase for the former, but only exhibits point singularities for the latter.  Importantly, while the localization is reduced, the topological invariant still exhibits a nontrivial value of $\pi$, while the $z$ values trace out similar trajectories to the PBC system.}

\section{Conclusion}

\textcolor{black}{We proposed and investigated two extensions of the generalized Bloch theorem based formalism of Alase et al.~\cite{Alase16,Alase17,Cobanera17,Cobanera18}.   On the one hand the original formalism was extended by a symmetry analysis; the bulk boundary correspondence indicator was calculated within a symmetry resolved basis of the symmetry operator that protects the topological zero modes.  This extension renders this quantity a true topological invariant.   We find that the topologically nontrivial phase is accompanied by a divergence in the bulk boundary correspondence indicator.  We also constructed an open boundary generalization of the Zak phase.  Our open boundary Zak phase calculations for the Kitaev model exhibit a nontrivial value of $\pi$ in the topologically nontrivial phase, and a value of zero in the trivial phase.} \\

We also applied our extensions to a system with a bond impurity.  In this case the bulk-boundary indicator was found to only be divergent at single points, at values of the chemical potential at which significant changes occur in the physical properties of the system.  A divergence occurs at the transition between the topologically trivial/non-trivial transition, and also, when the system enters the parameter region in which edge states become or cease to be oscillating states (known as the COO region).  In intermediate regions, the bulk boundary correspondence indicator is not divergent.   The open boundary Zak phase in this case behaved qualitatively the same way as for the open boundary Kitaev model.\\

We also studied how the parameters characterizing the $z$-transform evolve as the system is driven through the above transitions.  The positions of the $z$-values of the topological edge states on the complex plane are very informative.  This analysis is reminiscent of the study of Lee-Yang zeros~\cite{Lee52a,Lee52b,Timonin21,Chitov22}  in the classical Ising model.  Emphasis was also placed on the similarity of the formalism of Alase et al.~\cite{Alase16,Alase17,Cobanera17,Cobanera18} and the non-Bloch band theory developed~\cite{Yao18,Yokomizo19} to study non-Hermitian systems.  The edge state analysis based on the generalized Bloch theorem formalism is potentially useful in characterizing edge localization in non-Hermitian systems as well.  Another interesting possibility is to apply it in the case of the spectral bulk boundary correspondence~\cite{Tamura19,Daido19,Tamura21,Ahmed25} principle.\\

The central point of the modern polarization theory~\cite{King-Smith93,Resta94,Resta00,Vanderbilt18} is that in crystalline systems polarization can be expressed as a geometric phase~\cite{Berry84}, in this particular case, the Zak phase~\cite{Zak89}.  Topological invariants~\cite{Bernevig13,Asboth16} (such as the Majorana number~\cite{Kitaev01}) are modfied versions of the Zak phase, suited for the systems in question.  The modern polarization theory also provides tools to calculate moments and cumulants~\cite{Resta99,Souza00,Hetenyi19,Hetenyi22} associated with the polarization.  Our calculation of the Zak phase in an open boundary system opens the door to calculating moments and cumulants associated with topological invariants, possibly leading to a more complete characterization of such systems. \\

\section*{Acknowledgements}
This research was supported by HUN-REN 3410107 (HUN-REN-BME-BCE Quantum Technology Research Group), by the National Research, Development and Innovation Fund of Hungary within the Quantum Technology National Excellence Program (Project No. 2017-1.2.1-NKP-2017-00001), by Grants No. K142179, No. K142652, and No. FK142601 and by the BME-Nanotechnology FIKP Grant No. (BME FIKP-NAT).

\end{document}